\documentclass[prb,twocolumn,superscriptaddress,showpacs,floatfix]{revtex4}
\pdfoutput=1
\usepackage{amssymb,amsmath}
\usepackage{latexsym}
\usepackage{graphicx}
\usepackage{xspace}
\def\scoc{Sr$_2$CuO$_2$Cl$_2$\xspace}

\begin{document}

\title{Optically induced softening of the charge-transfer gap in \scoc}

\author{J.~S.~Dodge}
\affiliation{Department of Physics, Simon Fraser University, Burnaby, BC~V5A~16S, Canada}
\affiliation{Materials Sciences Division, E.~O.~Lawrence Berkeley
 National Laboratory, Berkeley, California 94720}
\author{A.~B.~Schumacher}
\altaffiliation[also: ]{~Institut f\"ur Angewandte Physik,
 Universit\"at Karlsruhe, 76128 Karlsruhe, Germany}
\affiliation{Materials Sciences Division, E.~O.~Lawrence Berkeley
 National Laboratory, Berkeley, California 94720}
\affiliation{Department of Physics, University of California at
 Berkeley, Berkeley, California 94720}
\author{L.~L.~Miller}
\affiliation{Ames Laboratory and Department of Physics, Iowa State
 University, Ames, IA 50011}
\author{D.~S.~Chemla}
\affiliation{Materials Sciences Division, E.~O.~Lawrence Berkeley
 National Laboratory, Berkeley, California 94720}
\affiliation{Department of Physics, University of California at
 Berkeley, Berkeley, California 94720}

\date{\today}

\begin{abstract}
Energy- and time-resolved spectroscopy reveals a photoinduced softening of the charge-transfer gap in the insulating copper oxide~\scoc that indicates rapid and efficient photoproduction of optical phonons. By relating the pump-probe signal amplitude to the thermal difference spectrum, we estimate that eleven to twenty optical phonons are created for every one 3~eV photon. Assuming relaxation to the optical absorption edge at 1.5~eV, this corresponds to 70--130~meV per boson. While the lower limit is consistent with relaxation exclusively through optical phonons, the upper limit suggests  a significant role for magnetic excitations. We observe a photoinduced bleaching of the gap excitation that we associate with phase space filling, and estimate the excluded area of the photoexcited state to be about nine copper oxide plaquettes. The temporal decay of the pump-probe signal is consistent with anharmonic phonon decay.
\end{abstract}

\pacs{71.27.+a,78.47.jc,71.38.-k,74.72.-h}

\maketitle

\section{\label{sec:intro}Introduction}
When a semiconductor is optically excited above the band gap, the carriers relax to the band edges by emitting a cascade of optical phonons on a picosecond timescale; the resulting nonequilibrium phonon population then relaxes by anharmonic processes over a few picoseconds.\cite{ref:von-der-Linde1988,ref:Shah1999} Time-resolved spectroscopy played an essential role in characterizing this process,\cite{ref:von-der-Linde1988,ref:Shah1999} and it is now being applied successfully in research on more complex solids.\cite{ref:Averitt2002,ref:Rini2007,ref:Hendry2007} The optical excitations of insulating magnetic oxides are an important frontier of this research. 

Understanding the elementary excitation spectrum of magnetic insulators has been a central concern in condensed matter physics for over half a century, and there is widespread consensus that high-temperature superconductivity is intimately related to the physics of the insulating, antiferromagnetic copper oxide plane.\cite{ref:Lee2006} The optical gap in lamellar insulating cuprates is associated with the transfer of an electron from an oxygen ion to a neighboring copper ion within the copper oxide plane.\cite{ref:Zaanen1985,ref:Tokura1990} The electron resides primarily on the copper in a nominal $3d^{10}$ state, while the oxygen hole forms a Zhang-Rice singlet with another neighboring copper in a $3d^{9}\underline{L}$ state.\cite{ref:Zhang1988} Beyond this basic picture, however, there continues to be controversy around the role played by interactions of the electron and hole with each other, with magnetic excitations, and with phonons.

The properties of the hole states in copper oxides have been extensively studied with angle resolved photoemission spectroscopy.\citep{ref:Damascelli2003} The hole bandwidth was measured to be $280\pm 60$~meV, strongly renormalized from the band theoretic value and consistent with the $2.2J$ prediction of the $t-J$ model, although an extension to a $t-t'-t''-J$ model is required to fit the detailed dispersion.\citep{ref:Wells1995,ref:Sushkov1997} The photoemission lineshape was observed to be unusually broad, and recent work has interpreted this as Franck-Condon broadening due to phonons that modulate the electronic and magnetic interactions through the Cu-O bond.\cite{ref:Shen2004,ref:Shen2007,ref:Mishchenko2004,ref:Rosch2005,ref:Sangiovanni2006}

The importance of carrier-phonon interactions to the optical properties of insulating cuprates were recognized by \textcite{ref:Falck1992} They showed that the energy and linewidth of the gap excitation have an anomalously strong temperature dependence, and explained this with a model involving short-range electron-hole interactions together with carrier-phonon interactions.\cite{ref:Falck1992} This behavior has since been explained alternately as the result of lattice expansion\cite{ref:Zibold1996} or electron-magnon interactions,\cite{ref:Choi1999} and currently the relative importance of these effects remains unresolved. Carrier-phonon interactions were further implicated in a study of the Urbach tail.\cite{ref:Lovenich2001}

Time-resolved optical studies in insulating cuprates have also provided evidence for strong coupling of carriers to phonons and magnons at lower energies, and recent work on photoinduced phase transitions have exploited this feature to produce metastably ordered states that would be impossible to create in thermal equilibrium.\cite{ref:Gedik2007,ref:Koshihara2008} Picosecond resonant Raman spectroscopy on insulating YBa$_2$Cu$_3$O$_{6.2}$ indicated that photoexcited carriers release excess energy to optical phonons in less than one picosecond.\cite{ref:Mertelj1997,ref:Poberaj1994} The photoexcited states in insulating YBa$_2$Cu$_3$O$_{6.2}$ and Nd$_2$CuO$_4$ have also been studied with transient absorption spectroscopy.\cite{ref:Matsuda1994} In that case, an observed sub-picosecond initial relaxation time was interpreted to be the rate at which photoexcited states give off energy to magnons.

In this work we study the optical gap excitation of the insulating cuprate \scoc with time- and energy-resolved laser spectroscopy, to probe the interaction between the optical gap excitation and bosonic modes at lower energies. We find that in response to photoexcitation, the energy gap softens and the optical spectral weight over the visible range is reduced. The photoinduced absorption spectrum mimics the effect of raising temperature, indicating that photoexcitation induces the same bosonic excitations that are excited thermally. By relating the pump-probe signal amplitude to the amplitude of the thermal difference spectrum we can infer that for every 3~eV photon absorbed, eleven to twenty low-energy bosons are emitted by the photoexcited state. For relaxation to the optical absorption edge at 1.5~eV, this corresponds to 70--130~meV per boson. While this range includes characteristic phonon energies, it extends well above them, indicating  a possible role for magnetic excitations in relaxing the energy of photoexcited carriers.

\section{\label{sec:method}Experimental Method}
All measurements were performed in a transmission geometry on high quality single crystals of \scoc that were cleaved to optically thin plates. \scoc stands out as an ideal model system for studying insulating cuprates, because the replacement of apical oxygen atoms by chlorine results in a very stable stoichiometry.\cite{ref:Miller1990,ref:Kastner1998} Several platelets were studied, all with consistent results. The measurements described here were performed on a $d=900$~\AA~thick \scoc single crystal platelet, held under vacuum in a cold finger cryostat to avoid surface degradation.

In the optical region, the dominant spectral feature of \scoc is the charge-transfer (CT) gap excitation at 2.1~eV.\cite{ref:Zibold1996,ref:Choi1999} We use 200~fs pulses of $h\nu_p = 3.1~\text{eV}$ light from a frequency-doubled regenerative amplifier  to create a non--equilibrium population of CT excitations. A white light continuum is used as a probe. We employ a zero dispersion stretcher to select a 7--9~nm wide probe pulse from the continuum that can be tuned from $h\nu = 1.6-2.3~\text{eV}$.

We determine the optical excitation density $N_\text{ex}$ using the expression
\begin{equation}
N_\text{ex} = \frac{E_0}{h\nu_p\pi w_{0h} w_{0v} d}(1-e^{-\alpha d}),
\label{eq:density}
\end{equation}
where $E_0$ is the laser pulse energy, $\nu_p$ is the pump laser frequency, $w_{0h}$ and $w_{0v}$ are the horizontal and vertical Gaussian beam waists, respectively, $d$ is the sample thickness, and $\alpha$ is the absorption coefficient at $\nu_p$. The Gaussian beam profiles of both the pump and the probe were characterized at the sample plane to an accuracy of $\pm1~\mu$m using the knife-edge method along both horizontal and vertical axes. During the course of our experiments the beam delivery optics varied, with a typical pump beam waist of $33~\mu$m. The pump beam was always 50\% larger than the probe beam to create a more homogeneous excitation profile for the probe. We estimate the total uncertainty in $N_\text{ex}$ to be 10\%.

At each center frequency, we measure the change in the transmitted probe intensity as the pump is chopped mechanically. In the limit of small signals, the normalized change in transmission reproduces the differential absorption of the sample, since $\Delta{\rm T/T} \equiv (I_t^{\prime}-I_t)/I_t \approx - \Delta \alpha \, d$, where $I_t^\prime$ and $I_t$ are the transmitted intensities with the pump on and off, respectively. During the measurement of $\Delta{\rm T/T}$ as a function of wavelength and time delay $\Delta t$, we reset the $\Delta t = 0$ position for each center wavelength. This is necessary to compensate for the dispersion experienced by probe pulses at different center frequencies.

\section{\label{sec:results}Results and Discussion}
\subsection{\label{sec:ppspec}Pump-probe spectrum}
\begin{figure}[t]
\begin{center}
 \includegraphics[width=0.95\columnwidth]{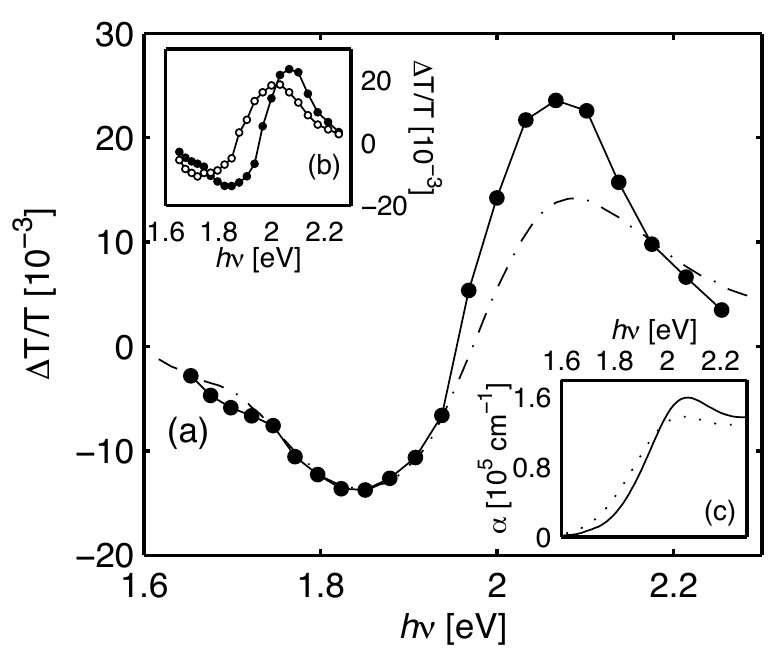}
 \caption{(a) Instantaneous ($\Delta t=50~\text{fs}$) photoinduced changes in \scoc after excitation with $h\nu_p=3.1~\text{eV}$ at $T=15~\text{K}$ ($\bullet$\ with solid line). The excitation density is $N_\text{ex} \approx 3.5\times10^{18}~\text{cm}^{-3}$, or $4\times 10^{-4}~\text{Cu}^{-1}$. Shown for comparison is the differential linear absorption, $-[\alpha(250~\text{K})-\alpha(15~\text{K})]d$, scaled by a factor of 0.07 (dash-dotted line). (b) Pump-probe spectrum at $\Delta t=50~\text{fs}$ for two temperatures, $T=15~\text{K}$ ($\bullet$) and $T=300~\text{K}$ ($\circ$). (c) Linear absorption of \scoc at $T=15~\text{K}$ (solid line) and $T=300~\text{K}$ (dotted line)}
 \label{fig:figure1}
\end{center}
\end{figure}

\begin{figure}[t]
\begin{center}
 \includegraphics[width=0.95\columnwidth]{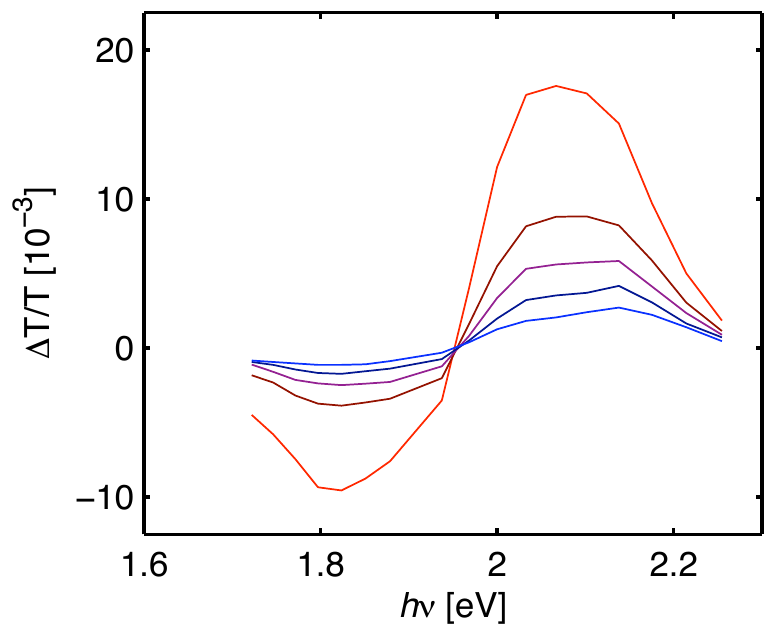}
 \caption{(Color online) Photoinduced absorption spectrum for $\Delta t = 1.7$~ps, 5.3~ps, 10~ps, 20~ps, and 50~ps, for $N_\text{ex} \approx 6.2\times 10^{18}\ \text{cm}^{-3}$. The signal amplitude decreases with time.}
 \label{fig:figure2}
\end{center}
\end{figure}

Fig.~\ref{fig:figure1}(a) shows the normalized change in transmission immediately following photoexcitation with $h\nu_p=3.1~\text{eV}$. 
We observe photoinduced absorption, $\Delta T<0$, at probe photon energies $h\nu<1.95~\text{eV}$, and  photobleaching, $\Delta T>0$, for  $h\nu > 1.95~\text{eV}$. In Fig.~\ref{fig:figure2}  we track the photoinduced absorption spectrum as a function of time delay, and we find that the transient spectral changes appear as one spectral unit instantaneously after the excitation, and decay---again as one spectral unit---on a picosecond timescale. In contrast to earlier work on similar cuprate compounds,\cite{ref:Matsuda1994} this leads us to conclude that the feature should be considered as a whole, rather than being due to two separate physical effects, one associated with absorption and the other with bleaching.

Viewed in this way, the effect of the pump beam is to broaden the gap excitation and shift it to lower energy. This interpretation is reinforced by comparing the pump-probe spectrum to the thermal difference spectrum $-[\alpha(T=250~K)$-$\alpha(T=15~K)]d$, also shown in Fig.~\ref{fig:figure1}. Apart from the stronger photobleaching in the pump-probe spectrum, the two curves show a remarkable similarity both in position and in overall spectral shape. Furthermore, as the temperature is raised from $T=15~\text{K}$ to $T=300~\text{K}$, the spectral shape of the pump-probe signal shown in Fig.~\ref{fig:figure1}(b) tracks the linear absorption spectrum shown in Fig.~\ref{fig:figure1}(c), broadening and shifting to lower energy.

These observations strongly suggest that the pump-probe response and thermally induced changes in the CT excitation result from the same underlying microscopic mechanism. Earlier studies found that both the energy and the linewidth of the CT excitation possess a temperature dependence that mimics a Bose-Einstein occupation function, with a characteristic energy scale of about 45~meV.\cite{ref:Falck1992,ref:Lovenich2001} Although this energy scale is characteristic of optical phonons in \scoc, we note that these can still couple strongly to the magnetic degrees of freedom by modulating the bonds through which magnetic exchange occurs.\cite{ref:Mishchenko2004,ref:Cataudella2007} Consequently we assign these low-energy bosons as optical phonons, with the understanding that they may have strong magnetic character.

\begin{figure}[t]
\begin{center}
	\includegraphics[width=0.95\columnwidth]{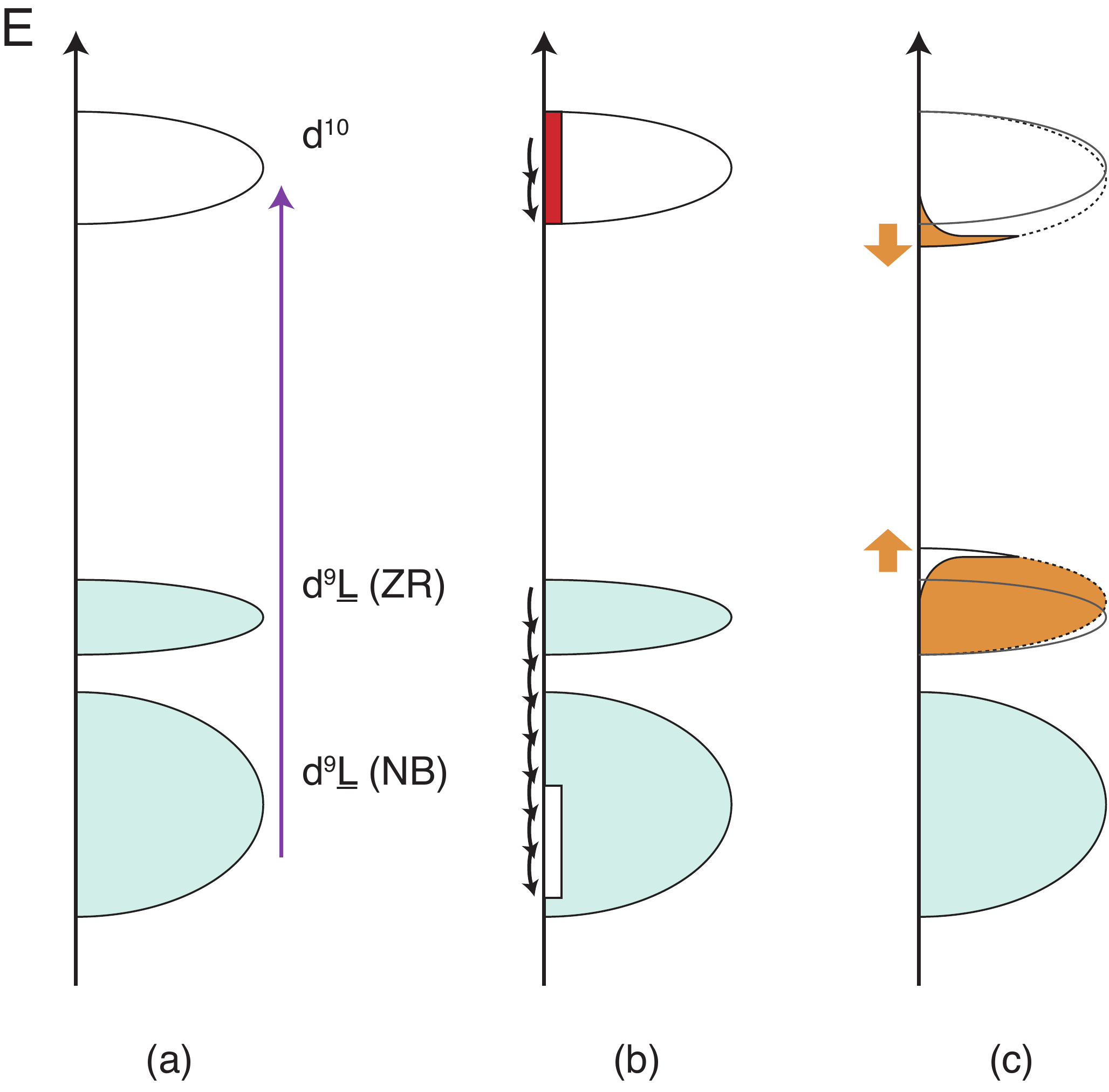}
\caption{(Color online) Model decay sequence for 3~eV optical excitation: (a) photon absorption at $t=0$; (b) rapid decay of hot carriers through phonon emission and carrier thermalization; (c) hot phonons cause the optical gap to broaden and shift.}
 \label{fig:figure3}
\end{center}
\end{figure}

We propose that photoexcitation triggers the same event sequence that is observed in conventional polar semiconductors,\cite{ref:von-der-Linde1988,ref:Shah1999} represented schematically in Fig.~\ref{fig:figure3}. Fig.~\ref{fig:figure3}(a) shows the initial photoexcitation with a 3~eV pump photon. These energetic CT excitations rapidly release excess energy via the emission of optical phonons, shown in Fig.~\ref{fig:figure3}(b). As we discuss below, there may be an additional relaxation channel involving magnetic excitations, but for simplicity we limit the discussion here to phonons. Once produced, the additional phonons have the same effect on the CT excitation as an increase in the equilibrium temperature of the lattice, shown in Fig.~\ref{fig:figure3}(c). In contrast with a simple heating effect, however, the photoinduced phonon population is strongly out of equilibrium, and will decay via anharmonic processes as the system thermalizes (not shown). The carrier distribution is shown thermalized in Fig.~\ref{fig:figure3}(c), as we expect this to occur rapidly during the phonon emission process.

The phonon emission time is inversely proportional to the electron-phonon interaction strength,\cite{ref:Kash1985,ref:Shah1999,ref:Conwell1967} given in the Fr\"olich model as
\begin{equation}
\alpha_F = \frac{e^2}{4\pi\hbar\epsilon_0}\left(\frac{1}{K_\infty} - \frac{1}{K_0}\right)\sqrt{\frac{m^*}{2\hbar\omega_0}},
\end{equation}
where $K_\infty$ and $K_0$ are the high-frequency and static dielectric constants, respectively, $m^*$ is the carrier effective mass, chosen to be $4m_e$, and $\omega_0$ is the optical phonon frequency. We take $\hbar\omega_0 = 45$~meV to be consistent with our earlier estimate,\cite{ref:Lovenich2001} although phonons with higher energies will tend to be favored as a way to release energy efficiently.\cite{ref:Shah1999} The experimental values for \scoc, $K_0 = 8.3$ and $K_\infty = 3.8$, give $\alpha_F \approx 5$. In conventional semiconductors the inelastic electron-phonon scattering rate can be derived perturbatively from the Fr\"olich model; in two dimensions, $1/\tau = \pi\alpha_F\omega_0$.\cite{ref:DasSarma1992} While \scoc is expected to be in the strong coupling limit, extending this perturbative result gives a crude estimate of $\tau\approx 1~\text{fs}$. The cascade following a 3~eV excitation could then generate 33 phonons as it relaxes to the gap edge at 1.5~eV, a process that would take about 33~fs to complete. Although this is still much less than the 200~fs resolution of our experiment, it is a timescale that is testable with current experimental capability.

With the results shown in Fig.~\ref{fig:figure1}(a) we can estimate the number of phonons $n_\text{\it eff}$ that are actually produced in the relaxation process by relating the amplitude of the pump-probe signal $\Delta T/T$ to the thermal difference spectrum $-\Delta\alpha\,d$,
\begin{equation}
n_\text{\it eff} = -\frac{1}{N_\text{\it ex}}\frac{\Delta T}{T}\frac{\Delta n_B}{\Delta\alpha\,d},
\label{eq:neff}
\end{equation}
where $\Delta n_B\approx 0.14$ is the change in boson occupation number in the thermal difference experiment, and $N_\text{\it ex} \approx 3.5\times 10^{18}\ \text{cm}^{-3}$ in the photoinduced absorption measurement. To determine $\Delta n_B$ we have again assumed a characteristic phonon energy of 45~meV.\cite{ref:Lovenich2001} The penetration depth of \scoc\ at 3~eV is 100~nm.\cite{ref:Zibold1996} As Fig.~\ref{fig:figure1}(a) shows, over the range 1.7--1.9~eV the thermal difference and pump-probe spectra are closely proportional; above $h\nu = 1.9$~eV the proportionality breaks down, although the spectral shape remains qualitatively similar. As we discuss further below, we assume that photoinduced bleaching is responsible for the difference at higher energies, and use the minimum of $\Delta T/T$ at $h\nu \approx 1.85$~eV as a proxy for the the phonon occupation level.

Combining these values in Eq.~\ref{eq:neff} gives $n_\text{\it eff} \approx 11$. If the carriers relax completely to the gap edge at 1.5~eV, this implies an average energy per boson of 130~meV, much larger than the 45~meV we assumed for the optical phonon energy. Alternatively, if we match the photoinduced absorption and thermal difference spectra at the peak near $h\nu\approx 2.1$~eV instead of at the dip near $h\nu\approx 1.85$~eV, we get $n_\text{\it eff}\approx 20$ and an average boson energy of 76~meV, comparable to the highest energy optical phonons in \scoc.\cite{ref:Zibold1996} Taking these as limiting cases, we get $n_\text{\it eff} \approx$~11--20.

This comparison suggests that magnetic excitations may play a role in relaxing the energy of the photoexcited state while contributing little to the pump-probe spectrum. One possible relaxation product is the resonant excitation of two magnons and one phonon,  which would absorb 350~meV of energy but create only one phonon.\cite{ref:Lorenzana1995,ref:Perkins1993} In this model, the average energy per boson can then be taken as a measure of the branching ratio between phonon relaxation and magnon relaxation. Taking the lower limit for $n_\text{\it eff}\approx 11$, a 3~eV optical excitation will release about 0.5~eV to phonons and 1.0~eV to magnetic excitations.

We note an additional weak photoinduced absorption feature at $h\nu=1.7~\text{eV}$ that we assign to the $\text{Cu}~3d_{x^2-y^2}\rightarrow 3d_{xy}$ ligand-field exciton.\cite{ref:Perkins1993,ref:Kuiper1998,ref:Choi1999,ref:Lovenich2001} This excitation is parity forbidden and appears as a weak shoulder extending from 1.4-1.7~eV in linear optical absorption measurements.\cite{ref:Perkins1993,ref:Choi1999,ref:Lovenich2001} It is evident in both linear absorption and in our pump-probe measurements at low temperatures only. Photoexcited charge carriers and optical phonons can both lead to a weak symmetry breaking that will induce absorption from this otherwise optically forbidden transition. 

\subsection{\label{sec:specwt}Spectral weight}
When integrated over the measured spectral range, the pump-probe spectrum shown in Fig.~\ref{fig:figure1}(a) exhibits a net photoinduced bleaching that is not observed in the thermal difference spectrum. Chemical doping causes dramatic changes to the spectral weight distribution of insulating cuprates,\cite{ref:Basov2005} and similar changes have been observed in photoexcitation experiments on timescales of microseconds and longer.\cite{ref:Kim1991,ref:Perkins1998} Such changes are commonly analyzed using the sum rule
\begin{equation}
\int_0^\infty\sigma(\omega)\,d\omega = \frac{\pi}{2}\frac{ne^2}{m},
\label{eq:sigmasum}
\end{equation}
where $n$ is the total electron number density and $m$ the electron mass. An equivalent sum rule can be written for the absorption coefficient $\alpha$ that is more convenient here:
\begin{equation}
\int_0^\infty\alpha(\omega)\,d\omega = \frac{\pi}{2c}\frac{ne^2}{m\epsilon_0}.
\label{eq:sigmasum2}
\end{equation}
These expressions indicate that the total frequency-dependent conductivity and absorption is conserved, so that any increase at one frequency must be compensated by a decrease elsewhere. In cuprates, increasing the carrier density causes spectral weight to be shifted from the charge-transfer gap to free-carrier conductivity at lower frequencies.

Photoinduced bleaching is well understood in semiconductor optics. It results from Pauli blocking, when photoexcited states occupy phase space that are then prevented by the Pauli exclusion principle from further population. To quantitatively assess the strength of the phase-space filling it is useful to calculate the real-space area that a photoexcited state excludes from further absorption\cite{ref:Chemla1984} To determine this for a 3~eV excitation in \scoc, we extract the amount of photobleaching $\Delta[\int\!\alpha(\omega)\,d\omega] \approx 230~\text{eV\,cm}^{-1}$ in the pump-probe response at $T=15~\text{K}$ and normalize it by the total linear absorption $\int\!\alpha(\omega)\,d\omega\approx 6.3\times 10^4~\text{eV\,cm}^{-1}$ over the same spectral range to get a fractional absorption of $3.7\times 10^{-3}$. Dividing this by the excitation density $4\times 10^{-4}~\text{Cu}^{-1}$ gives excluded area of 9.3 copper oxide plaquettes. This number suggests that the photoexcited state is distributed over a several lattice spacings, as would be expected from a hole state that is participating in a Zhang-Rice singlet.

\subsection{\label{sec:tdep}Time dependence}
\begin{figure}[t]
\begin{center}
	\includegraphics[width=0.95\columnwidth]{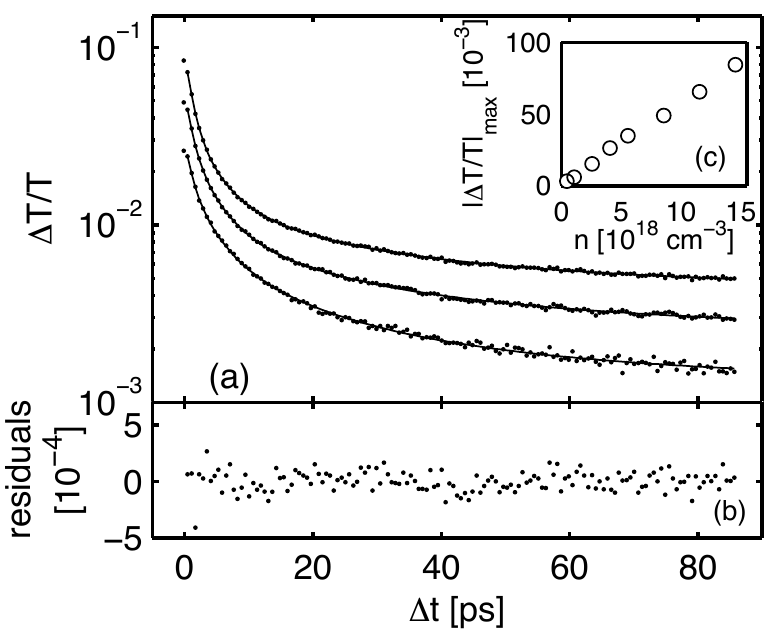}
	\caption{(a) Decay dynamics observed at $T = 15~\text{K}$ for $h\nu_p=3.1~\text{eV}$, probe energy $h\nu=2.1~\text{eV}$, and three different excitation densities ($4, 9, 15\times 10^{18}~\text{cm}^{-3}$ from bottom to top). Measurements ($\cdot$) are shown together with fits (thin lines) to Eq.~\ref{eq:decay}. (b) Residuals of the fit for $n = 2.4\times 10^{19}~\text{cm}^{-3}$. (c) Power dependence of the instantaneous signal as a function of the excitation density $n$.}
 \label{fig:figure4}
\end{center}
\end{figure}

\begin{figure}[t]
\begin{center}
	\includegraphics[width=0.95\columnwidth]{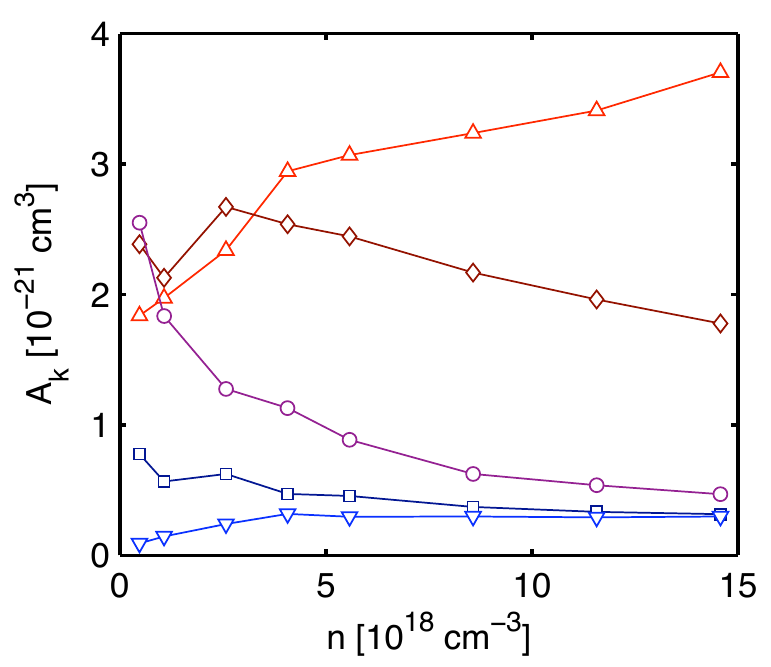}
\caption{(Color online) Fit amplitudes $A_1$ (red $\vartriangle$), $A_2$ (dark red $\Diamond$), $A_3$ (purple $\circ$), $A_4$ (dark blue $\Box$), and  $A_\infty$ (blue $\triangledown$) as a function of density $n$, with optimized values $\tau_1 = 1.1~\text{ps}, \tau_2 = 3.7~\text{ps}, \tau_3 = 12.6~\text{ps}, \text{and } \tau_4 = 42.7~\text{ps}$. }
 \label{fig:figure5}
\end{center}
\end{figure}

If the pump-probe response is generated by phonons as we have argued, we expect its temporal behavior to be determined by phonon decay mechanisms. Fig.~\ref{fig:figure4}(a) shows the decay of the pump-probe response with 
$h\nu_p=3.1~\text{eV}$, $h\nu=2.1~\text{eV}$ and $T=15~\text{K}$ at three different excitation densities. The decay dynamics depend weakly on density and are strongly nonexponential. They are not well described by stretched exponential fits or bimolecular decay kinetics, so to facilitate quantitative comparison of the dynamics at different densities, we employ a discretized Laplace transform, fitting the data by a sum of exponential decays:
\begin{equation}
\frac{\Delta T(\Delta t>0)}{nT} = A_\infty + \sum_{k=1}^N A_k e^{-\Delta t/\tau_k}.
\label{eq:decay}
\end{equation}
The decay times $\tau_k$ are globally constrained to a logarithmic spacing $\tau_k = \lambda^{k-1}\tau_1$, yielding constrained linear fits to $\{A_1,\ldots,A_N,A_\infty\}\geq 0$ that may be compared directly as a function of excitation density.\cite{ref:Istratov1999} We find that $N=4$ is the minimum value necessary to achieve a good fit, with residuals as shown in Fig.~\ref{fig:figure4}(b) for the highest density. Fig.~\ref{fig:figure4}(c) shows that up to the highest pump intensities available to us, the photoinduced signal is linear in the excitation density. Consequently, we expect $A_\infty + \sum_{k=1}^NA_k= C$, and an increase in one amplitude will come at the expense of others. The resulting amplitudes are shown as a function of density in Fig.~\ref{fig:figure5}.

The picosecond decay timescale is typical for anharmonic decay of LO phonons into two LA phonons at lower energies,\cite{ref:von-der-Linde1988} and given the complexity of the \scoc crystal structure the distribution of decay timescales is unsurprising. As the density increases, the amplitudes are shifted to shorter timescales, indicating an increase in the number of decay channels as the phonon density increases. We find relatively small temperature dependence of the decay dynamics up to room temperature, expected for optical phonons that will decay primarily through spontaneous emission. After the charge carriers have relaxed to the minimum gap energy of 1.5~eV, we expect them to dissipate their remaining energy through either radiative or nonradiative recombination on longer timescales.

\section{\label{sec:sum}Summary}
In summary, we have characterized the photoinduced absorption spectrum of \scoc as a function of excitation density, temperature, and pump-probe time delay, and found that the characteristic features of the response can be explained in terms of a strong coupling between the CT excitation and optical phonons. We have estimated that from eleven to twenty phonons are created upon photoexcitation at 3~eV. A value at the lower bound would account for only a third of the total energy that the carriers must release to relax to the gap edge, and we have argued that magnetic excitations may also play a role in the relaxation process. We have observed a photoinduced bleaching of the charge-transfer excitation and used this to estimate the size of a 3~eV excitation to be about nine CuO$_2$ plaquettes. Finally, we have characterized the temporal decay of the photoexcited state and associated it with anharmonic phonon decay. Our experiments suggest a novel way to use light to manipulate the low-energy bosonic excitations of insulating cuprates and probe their interactions with other optical excitations. 

\acknowledgements
This work was supported by the Director, Office of Science, Office of Basic Energy Sciences, Division of Materials Sciences and Office of Science, U.S. Department of Energy, under Contracts No. DE-AC03-76SF00098 and W-7405-Eng-82. JSD acknowledges support from the Natural Science and Engineering Research Council of Canada (NSERC) and the Canadian Institute for Advanced Research (CIFAR). ABS gratefully acknowledges a fellowship by the German National Merit Foundation.


\end{document}